\documentclass{llncs}

\usepackage{times}
\usepackage{scriptcode}

\title{An Approach to the Implementation of Overlapping Rules in Standard ML}
\author{Riccardo Pucella}
\institute{Department of Computer Science\\Cornell University\\riccardo@cs.cornell.edu}

\newcommand{\COMMENTOUT}[1]{}
\newcommand\kw[1]{\textbf{#1}}
\newcommand\expr[1]{\textit{#1}}

\begin{document}
\maketitle

\begin{abstract}
We describe an approach to programming rule-based systems in Standard
ML, with a focus on so-called overlapping rules, that is rules that
can still be active when other rules are fired. Such rules are useful
when implementing rule-based reactive systems, and to that effect we
show a simple implementation of Loyall's Active Behavior Trees, used
to control goal-directed agents in the Oz virtual environment. We
discuss an implementation of our framework using a reactive library
geared towards implementing those kind of systems.
\end{abstract}

\section{Introduction}

Rule-based systems\footnote{In this paper, we focus exclusively on
production rule systems. Related systems, such as those based on a
notion of term rewrite rule, have not been considered at this point.}
have had a long history in AI and powerful 
implementations have been developed.  The
most problematic aspect of this work 
has always been that of integrating the rule-based approach to a
general-purpose language for application support. As an example along
those lines, in \cite{Crawford96}, Crawford \emph{et al} describe R++,
a rule-based extension to C++ completely integrated with the
object-oriented features of the language, implemented as a rewrite of
R++ into C++ code.  

\COMMENTOUT{
Our aim in this paper is somewhat similar, but along different
lines. We attempt to integrate rule-based programming with Standard ML 
(SML) \cite{Milner97}, in such a way as to be able to use both
rule-based code and ordinary SML code. Our integration should be as
much a library as possible, that is it should use facilities of the
}

One common use of rules, and indeed the primary motivation for this
work, is to help write rule-based reactive systems, reactive in the sense of having the
system react to changes in the environment. A change in the
environment should enable a certain number of appropriate rules that
can be fired to react to the environment change, possibly effecting
new changes to the environment that will enable other rules. 

\COMMENTOUT{With this 
point of view in mind, it is not surprising that the approach we
investigate relies on various reactive programming approaches. More
precisely, we investigate the use of a general reactive library
introduced in \cite{Pucella98} and based on the Reactive Approach of
Boussinot \cite{Boussinot91,Boussinot96b}. }

One aspect of rule-based reactive systems we focus on is that of
long-acting overlapping rules. In 
most rule-based systems, a single rule is active at any given time:
when the system determines that a rule is enable, the rule is fired,
the environment is updated, and a new rule can be selected. This is
not so great is some of the rules are computationally intensive: when
such a rule is fired, it will take time to execute and if the system
relies on other rules to ensure say responsiveness of the interface,
the system will not respond to the user until the rule has finished
executing. In general one, may want rules that span multiple other
rules firing. For example, one may pre-fire a rule when certain
conditions are met, perhaps pre-computing some values, and get ready
for the ``real'' firing once the ``real'' conditions are in place. Or
one may want a rule that when fired will perform a given action at
every subsequent rule firing until maybe a condition occurs that stops 
this behavior. It is of course possible to achieve these effects in
certain rule-based systems, by a process of chaining rules (at the end 
of one rule, setting up the firing conditions of the next rule), along
with a suitable notion of concurrently fired rules. 

In this paper, we describe an approach to the integration of
rule-based programming in Standard ML (SML) \cite{Milner97}, with a
strong focus on overlapping rules to achieve the effects described
above. The resulting framework is suitable for the design of
domain-specific abstractions for various rule systems. In contrast with 
other work, we do not worry about efficiency issues in this paper, but 
rather concentrate on expressiveness and applicability of the
framework. We assume throughout the paper a basic knowledge of SML, as 
described in various introductory material such as
\cite{Paulson96,Ullman98}. 

After reviewing the basic notions of rule-based programming in Section
\ref{s:review}, we discuss the framework in Sections
\ref{s:basic}--\ref{s:overlapping}. We give an application of the
flexibility of the framework in Section \ref{s:abts}, where we design
domain-specific abstractions for controlling a goal-directed
agent. Section \ref{s:impl} focuses on implementation details,
including a reference implementation in terms of an existing reactive
library for programming reactive systems in SML, introduced in
\cite{Pucella98}, and based on the reactive approach of Boussinot
\cite{Boussinot91,Boussinot96b}.

\section{OPS-style production rules}
\label{s:review}

In this section, we establish the terminology and the model of rules
we are interested in, namely OPS-style production rules
\cite{Brownston85}. The \emph{production-system} model of  
computation is a paradigm on the same footing as the procedural
paradigm, the functional paradigm or the object-oriented paradigm: it
is a view of what a computation ought to be to best achieve a given
goal. 

A \emph{production-system program} is an unordered collection of basic 
units of computation called \emph{production rules} (henceforth simply 
called rules). Each rule has a condition part and an action part. An
\emph{inference engine} is used to execute the rules: it determines
which rules are \emph{enabled} by checking which conditions are true,
and select rules to execute or \emph{fire} from the enable rules. A
rule is fired by executing its action part, which typically will have
a side effect of performing input and output or computing a value and
updating some date in memory. Thus rule-based systems fundamentally
based on the notion of side effects.

Many mechanisms can be used to select which enabled rule to fire. In
the literature, the term \emph{conflict set} is often used to name
the set of rules which are enabled, reflecting the intuition that
somehow these rules can be conflicting, that is update the store in
different incompatible ways. To ensure such problems do not occur,
production systems will typically select a single rule to fire, by
methods involving various notions such as priority or probabilities,
along with notions such as the best matching of the conditions and so
on. 

In a rule-based system, control is \emph{data-driven}, that is the
data determines which part of the program will execute ---
furthermore, communication between different units is done solely
through the use of data. There is no concept of a subroutine call to
another rule, or anything of that sort. Rule-based systems allow a
cleaner separation of knowledge (in the form of rules) from the control
(encapsulated in the inference engine). This makes rule-based systems
well-suited to program expert systems for analysis problems, and for
programs for which the exact flow of control is not known. In this
paper, we will make the further refined point that with the
appropriate extensions, rule-based systems are well-suited for the
compositional development of reactive systems.

\section{A framework for rule-based programming}
\label{s:basic}

We begin by considering a general framework for the handling of simple OPS-style
rules in SML, where actions are executed atomically and terminate
before a new rule can be fired. We discuss the framework abstractly,
that is in terms its interface. 

The first notion of importance is that of a set of rules, as an
abstract type with a single basic operation that creates an empty set
of rules. The reason for keeping the set of rules abstract is to
allow for different implementations, some possibly aimed at optimizing 
the evaluation of the conditions.
\begin{code}
  \kw{type} rule_set
  \kw{val} newSet : unit -> rule_set
\end{code}

A rule is simply defined as a pair of a condition and an action, as in 
OPS.  Since we want the condition to be dynamically
evaluated, it is implemented as a function (the standard way to delay
evaluation in an eager language such as SML). A
condition evaluates to a positive integer (a word), which we call the
\emph{fitness} of the condition, indicating the degree to which the
condition is satisfied. There is no \emph{a priori} semantics or range 
associated with those, they are left to the discretion of the
programmer. The only restriction is that a fitness of zero is used to
indicate that a rule is not enabled. One can implement simple boolean conditions as values 0
and 1, if need be. \COMMENTOUT{A more involved implementation would push the
fitness into an abstract type with an associated ordering, which is
what we really require from words.} The only operation on rules is to
add them to rule sets. Note that because functions are first-class in
SML, rules become first-class as well. That is, they can be passed as
argument to functions, and returned from functions. 
\begin{code}
  \kw{type} rule = \{cond : unit -> word,
               action : unit -> unit\}
  \kw{val} addRule : rule * rule_set -> rule_set
  \kw{val} mkSet : rule list -> rule_set
\end{code}

The final ingredient of the system is the inference
engine, that selects which of the rules will be fired. At this point, the
issue of \emph{when} the rules should be fired must be
addressed. Many systems tie the firing of the rules, or at least the
evaluation of the conditions, to a change to variables that affect
the conditions. We choose a much more fundamental approach using an
explicit call that monitors the conditions, from which we can derive a
\emph{change-of-state} trigger. While inefficient, this approach has
the advantage of being general. The other issue this does
not address is that of conflict resolution, which rules to fire
of all those that are enabled. We provide a selection of conflict
resolution strategies. A function \expr{monitor} is used to select
rules to fire in a given set of rules. It takes as argument a conflict 
resolution flag determining the resolution strategy to use:

\begin{code}
  \kw{datatype} conflict_res = AllBest
                        | RandBest
                        | AllDownTo \kw{of} word 
                        | RandDownTo \kw{of} word
  \kw{val} monitor : conflict_res -> rule_set -> unit
\end{code}
\expr{AllBest} fires all the rules that qualify as the best rule to
apply, sorted according to fitness. \expr{RandBest} randomly picks one
of the rules that qualify as the best rule. \expr{AllDownTo} and
\expr{RandDownTo} perform similarly, but consider all the rules whose 
fitness is at least the given value. One can easily extend
the framework to allow for custom conflict resolution, which we do not
pursue in this paper for simplicity. 

Our notion of fitness is general. As we noted,
we can imagine a binary fitness (0-1) for boolean firing, but also a
fitness based on how close the conditions are to being completely
satisfied, or even so far as how many conditions are actually
satisfied (if we allow firing based on partially satisfied
conditions). An easy extension to the 
framework would be to pass information from the condition to the
action. We can mimic this easily by using a reference cell which can
also be hidden in a closure of the rule, as follows:
\begin{code}
  \kw{let} 
    \kw{val} r = ref 0
  \kw{in}
    \{cond = \textit{compute fitness, store something in r},
     action = \textit{some action using the value in r}\}
  \kw{end}
\end{code}

In striking difference with other systems, rules are not persistent in 
this framework: once a rule fires and executes, it is removed from the 
rule set. To make a rule persistent, we can use the function
\expr{persistent}. This allows us to dispense
with a function to remove rules 
from rule sets. Moreover, \expr{persistent} comes for
free given our extension for managing overlapping rules, as we will
see in the next section.
\begin{code}
  \kw{val} persistent : rule -> rule
\end{code}

As an example of how to use the framework, consider a simple
rule-based program to compute the greatest common divisor of two
integers, the classic Euclid's algorithm. The example is artificial
(it is easily implemented in SML without rules), but serves well to
illustrate the basics. A more complete example is presented in Section 
\ref{s:abts}. The program can be expressed
as follows in Dijkstra's language of guarded commands:
\[ \kw{do}~ X > Y \rightarrow X := X-Y ~|~ Y > X \rightarrow Y := Y - X ~\kw{od}\]
The corresponding code in our framework is more verbose, but
essentially similar:
\begin{code}
  \kw{fun} gcd (x,y) = \kw{let}
    \kw{val} rx = ref x
    \kw{val} ry = ref y
    \kw{fun} r (r1,r2) = persistent 
                      \{cond = \kw{fn} () => \kw{if} (!r1) > (!r2) 
                                         \kw{then} 1 \kw{else} 0,
                       action = \kw{fn} () => r1 := (!r1) - (!r2)\}
    \kw{val} rs = mkSet [r (rx,ry),r (ry,rx)]
    \kw{fun} loop () = \kw{if} ((!rx) = (!ry)) \kw{then} (!rx) 
                    \kw{else} (monitor AllBest rs; loop ())
  \kw{in}
    loop ()
  \kw{end}
\end{code}

The above example is interesting because it shows how the fact that
rules are first-class in the framework allow for parametrized rules:
the rule \expr{r} in the above code is parametrized over the reference 
cells containing the two arguments, parameterization which nicely
showcases the symmetry of the rules.

As a final remark, we note that the rules as we have presented them in
this section are simpler than they are in the actual framework. The rules
in the implemented framework contain an extra field generically called
\emph{data} whose type is a parameter to the structure implementing
the rules (in effect, the library is implemented as a functor). The
monitoring function takes as an extra argument a function to compute a 
fitness both from the result of the evaluation of the condition and
the data (which for example can contain notions such as rule priority
and so on). Since describing this explicitly would require us to go
into the details of both the module system and the type system of SML, we punt
on these issues in this paper. 

\COMMENTOUT{
One advantage of our approach is that we can fairly easily derive new
rule representations with hard-wired behavior. For example, assume we
want to implement prioritized rules. We can create a new structure
\begin{code}
  \kw{structure} PriorityRules :> \kw{sig}
    \kw{type} rule_set
    \kw{val} new_set
    \kw{type} rule = \{fitness : unit -> word,
                 action : unit -> unit,
                 priority : unit -> word\}
    \kw{val} addRule : rule_set * rule -> rule_set
    \kw{val} monitor : rule_set -> rule_set
  \kw{end} = \kw{struct}
    \kw{structure} PR = RulesFn (\kw{type} rules_data = unit -> word)
    \kw{type} rule_set = PR.rule_set
    \kw{val} new_set = PR.new_set
    \kw{type} rule = \{fitness : unit -> word,
                 action : unit -> unit,
                 priority : unit -> word\}
    \kw{fun} add_rule rs \{fitness,action,priority\} = PR.add_rule rs \{fitness=fitness,action=action,data=priority\}
    \kw{val} monitor = PR.monitor (\kw{fn} \{fitness,data\} => (fitness() * data ()),PR.RandBest)
  \kw{end}
\end{code}
}

\section{Managing overlapping rules}
\label{s:overlapping}

The main point of this work was to introduce \emph{overlapping rules}, 
that is rules that can span multiple invocations and be performed in 
parallel with other rules. From an interface point of view, the
framework only requires the addition of a single primitive, which we
call \expr{wait}, to add the desired functionality:
\begin{code}
  \kw{val} wait : (unit -> word) option\footnote{a primitive SML type defined as \cd{\kw{datatype} 'a option = NONE | SOME \kw{of} 'a}.} -> unit
\end{code}

The semantics of \expr{wait} is simple. Fundamentally,
\expr{wait} interrupts the action of the current rule, as if the rule
was finished executing, except that at the next time the monitoring
function is invoked to fire a rule, the rules that were interrupted are
allowed to resume while the new rule fires. Hence the term
\emph{overlapping}. In fact, the \expr{wait} primitive takes two
forms. The form \expr{wait (NONE)} behaves as an unconditional
interruption. Execution continues the next time the monitoring
function is invoked. The form \expr{wait (SOME (f))} with \expr{f} as
a condition (that is, a \expr{unit $\rightarrow$ word} function) 
also interrupts, but only resumes the rule the next time the
monitoring function is invoked with the condition \expr{f} being
satisfied. In effect, \expr{wait (SOME (f))} interrupts the rule and
conceptually replaces it with a new rule containing the remainder of
the interrupted rule, with a condition \expr{f}. 

As we mentioned in the previous section, rules are not persistent:
once they are fired and execute, they are removed from the rule
set.We can implement the \expr{persistent} function using \expr{wait}: 
\begin{code}
  \kw{fun} persistent \{cond,action\} = 
    \{cond = cond,
     action = \kw{let}
                \kw{fun} loop () = (action (); 
                               wait (SOME (cond));
                               loop ())
              \kw{in}
                loop
              \kw{end}\}
\end{code}
This nicely shows the power of first-class rules.

\section{An application: goal-directed agent control}
\label{s:abts}

We describe in this section one of the motivating applications for the 
development of the framework in the first place, that of controlling
goal-directed agents. The architecture we have in mind is inspired by
Hap, a reactive, goal-driven architecture for controlling agents in
the Oz virtual environment \cite{Loyall91}. The main structure in Hap
is an \emph{active behavior tree} (ABT), which represents all the
goals and behaviors an agent is pursuing at any given point
\cite{Loyall97}. An agent chooses the next step to perform by selecting
one of the leaves of its ABT. Three types of actions can be performed
depending on the type of the node selected. 
\begin{enumerate}
\item \textbf{Primitive physical action}: an action sent to 
the action server, which can either succeed or fail depending on the
state of the world.
\item \textbf{Primitive mental action}: an action that simply performs
a computation (possibly with side effects) and which always succeeds.
\item \textbf{Subgoal}: an action corresponding to a new subgoal; an
appropriate behavior is selected that matches that subgoal, and the
ABT is expanded by adding the steps specified by the behavior to the
tree as children of the subgoal selected. 
\end{enumerate}

(For our purposes, we drop the distinction between mental and physical 
actions, since we can model mental actions as physical actions that
always succeed). Programming an agent reduces to programming behaviors
for various goals. Goals have no intrinsic meaning, they are simply
names on top of which the programmer can 
attach any semantics she desires. Behaviors are defined by specifying
to which goal they apply, a pre-condition for the application of that
behavior (simply a predicate over the state of the world), and the
steps that the behavior prescribes (physical actions, mental actions,
subgoals). At this point, many details enter the description, to
provide control over managing goals and behaviors. There are three kind
of behaviors:
\begin{enumerate}
\item \textbf{Sequential}: the steps are performed in order, and
failure of any step signifies the failure of the corresponding goal to
which the behavior is attached; success of the last step signifies
success of the corresponding goal.
\item \textbf{Concurrent}: the steps are performed in any order, but
again failure of any step signifies the failure of the corresponding
goal.
\item \textbf{Collection}: the steps are performed in any order, but
success or failure of the steps are irrelevant. When all steps have
succeeded or failed, the corresponding goal succeeds. 
\end{enumerate}
Subgoals steps in behaviors can moreover be annotated as \emph{persistent},
that is when they succeed or fail, they are not removed, but rather
persist as a continuing goal. Typically, top-level goals are
persistent. Conversely, subgoals can be annotated with a \emph{success
test}, a predicate over the state of the world, which gets tested
every time the ABT is activated. If the success test of a subgoal
is true, the subgoal automatically succeeds. 

Choosing a step to perform is by default done at random over all the
applicable steps in an ABT, that is all the leaves that can be either
executed right away or subgoals that can be expanded because a
behavior applies (no behavior may apply because either none has been
defined or no pre-condition is satisfied). Similarly, choosing a
behavior to perform once a subgoal step has been chosen is by default
done at random over all applicable behaviors. One can modify this
default by assigning \emph{priorities} to various subgoals. 

All of this is meant to evoke the kind of structure we would like to
express in our framework. Since Hap revolves around the notion
of goals, we abstractly provide a notion of goal to the framework of
the previous sections, where goals are for simplicity represented as
strings.  
\begin{code}
  \kw{datatype} goal_status = Success | Failure 
                       | Active | Available 
                       | NoSuch
  \kw{val} goalSet : string -> unit
  \kw{val} goalSucceed : string -> unit
  \kw{val} goalFail : string -> unit
  \kw{val} goalStatus : string -> goal_status
  \kw{val} goalClear : string -> unit
\end{code}
where \expr{goalSet} enables the given goal, such that it is
to be pursued by the agent (it becomes available). The functions
\expr{goalSucceed} and \expr{goalFail} are used to record that a goal
has succeeded or failed. The function \expr{goalStatus} returns the
status of the given goal. The status of a goal is either
\expr{Success} or \expr{Failure} if the goal has been recorded as
such, or \expr{Active} if a behavior is actively pursuing the goal,
but is not done with it yet. A status of \expr{Available} indicates
that the goal is enabled, but that no behavior is pursuing it, while a 
status of \expr{NoSuch} indicates that no such goal exists. The
function \expr{goalClear} removes a goal from the active  
list of goals. We define the boolean-valued helper functions
\expr{isAvailable} and  \expr{isDone} to check the status
of a goal to be respectively \expr{Available} or
\expr{Success}/\expr{Failure}. 

We interpret an ABT behavior as a rule, triggered both by the
pre-condition of the behavior (if present) and the apparition as
\expr{Available} of the goal the behavior is meant to pursue. We do
not worry about either sequential, collection or concurrent
annotations, choosing rather to let the programmer manage the steps of 
the behavior explicitly. Patterns quickly emerge. For instance, a behavior for goal 
\expr{g} triggered by a condition \expr{c} and sequentially performing 
subgoal \expr{g1}, action \expr{a} and subgoal \expr{g2} can be
interpreted as a rule:
\begin{code}
  \kw{val} beh1 = 
   \{cond = \kw{fn} () => \kw{if} isAvailable (\expr{g}) \kw{andalso} \expr{c} 
                      \kw{then} 1 \kw{else} 0,
    action = \kw{fn} () => 
               (goalSet (\expr{g1});
                wait (SOME (\kw{fn} () => isDone (\expr{g1})));
                \kw{case} (goalStatus  (\expr{g1}))
                  \kw{of} Success => (goalClear (\expr{g1});
                                 \expr{a};
                                 wait (NONE);
                                 goalSet (\expr{g2});
                                 wait (SOME (\kw{fn} () => isDone(\expr{g2})));
                                 \kw{case} (goalStatus (\expr{g2}))
                                   \kw{of} Success => (goalClear (\expr{g2});
                                                  goalSucceed (\expr{g}))
                                    | _ => (goalClear (\expr{g2});
                                            goalFail (\expr{g})))
                   | _ => (goalClear (\expr{g1});
                           goalFail (\expr{g})))\}
\end{code}

Similarly, the previous behavior can be implemented concurrently by
setting all the goals at once and waiting for all the goals to be done. 
\begin{code}
  \kw{val} beh2 = 
   \{cond = \kw{fn} () => \kw{if} isAvailable (\expr{g}) \kw{andalso} \expr{c} 
                      \kw{then} 1 \kw{else} 0,
    action = \kw{fn} () => 
               (goalSet (\expr{g1});
                goalSet (\expr{g2});
                \expr{a};
                wait (SOME (\kw{fn} () => isDone (\expr{g1}) \kw{andalso} isDone (\expr{g2})));
                \kw{case} (goalStatus (\expr{g1}),goalStatus (\expr{g2}))
                  \kw{of} (Success,Success) => (goalClear (\expr{g1});
                                           goalClear (\expr{g2});
                                           goalSucceed (\expr{g}))
                   | (_,_) => (goalClear (\expr{g1});
                               goalClear (\expr{g2});
                               goalFail (\expr{g})))\}
\end{code}

By virtue of the \expr{andalso} in the first waiting condition of
\expr{beh2}, the conditions must all be true for the system to proceed 
at that point. Although it does sequentializes the testing of the
conditions, this is not an issue given our current framework since
the rule is resumed when all the conditions are satisfied. It may
however become an issue if we attempt to optimize the satisfaction of
the conditions. Persistence of goals can be implemented by
clearing them and setting them again, while goal success tests can be
wrapped inside the \expr{wait} condition for that goal.

One difficulty of this approach, immediately noticeable from the above 
code, is
that it is very error-prone, even if it is much more
flexible than ABTs. In effect, we have to implement the handling
of the goals explicitly, for every single behavior. However, the
flexibility of
first-class rules and first-class functions allows us to easily
generate such rules from a declarative description of the intended
behavior. Consider a type \expr{behavior} that describes a behavior
declaratively: 
\begin{code}
  \kw{type} behavior = \{goal : string,
                   precond : (unit -> bool) option,
                   kind : behavior_kind,
                   steps : behavior_step list\}

  \kw{datatype} behavior_kind = Sequential | Concurrent
  \kw{datatype} behavior_step = Subgoal \kw{of} string
                         | Action \kw{of} unit -> bool
\end{code}
(For simplicity, we drop the collective kind of behavior, its handling 
similar enough to the concurrent one to not cause a problem). We can
describe the previous two behaviors \expr{beh1} and \expr{beh2} as
follow (this time using behavior parameterization!):
\begin{code}
  \kw{fun} beh (k) = \{goal = \expr{g},
                 precond = SOME (\kw{fn} () => \expr{c}),
                 kind = k,
                 steps = [Subgoal (\expr{g1}),
                          Action (\kw{fn} () => \expr{a}),
                          Subgoal (\expr{g2})]\}
  \kw{val} beh1 = beh (Sequential)
  \kw{val} beh2 = beh (Concurrent)
\end{code}

We can then interpret such descriptions in our framework, by a
function \expr{behaviorRule}, which takes a description of type
\expr{behavior} and returns a rule of type \expr{rule}. The
implementation of \expr{behaviorRule} is simply a question of writing
a rule whose action is an interpreter for lists of
\expr{behavior\_step}. For the truly interested, the code for
\expr{behaviorRule} is given in Figure \ref{f:behaviorrule}.

\begin{figure}
\hrule
\medskip
\begin{code}
  \kw{fun} behaviorRule \{goal,precond,kind,steps\} = \kw{let}
    \kw{fun} split [] = ([],[])
      | split (x::xs) = \kw{let}
          \kw{val} (sg,acts) = split (xs)
        \kw{in}
          \kw{case} x
            \kw{of} Subgoal (g) => (g::sg,acts)
             | Action (a) => (sg,a::acts)
        \kw{end}
    \kw{fun} cond () = \kw{if} isAvailable (goal) \kw{andalso}
                     (\kw{case} precond \kw{of} NONE => true
                                    | SOME (pc) => pc ())
                         \kw{then} 1 \kw{else} 0
  \kw{in}
    \kw{case} kind
      \kw{of} Sequential => \kw{let}
           \kw{fun} perform_steps [] = goalSucceed (goal)
             | perform_steps (Action (a)::r) = 
                      (a ();
                       wait (NONE);
                       perform_steps (r))
             | perform_steps (Subgoal (g)::r) = 
                      (goalSet (g);
                       wait (SOME (\kw{fn} () => isDone (g)));
                       \kw{case} (goalStatus (g))
                         \kw{of} Success => (goalClear (g);
                                        perform_steps (r))
                          | _ => (goalClear (g);
                                  goalFail (goal)))
         \kw{in}
           \{cond = cond,
            action = \kw{fn} () => perform_steps (steps)\}
         \kw{end}
       | Concurrent => 
           \{cond = cond,
            action = \kw{fn} () => \kw{let}
              \kw{val} (subgoals,actions) = split (steps)
            \kw{in}
              app goalSet subgoals;
              app (\kw{fn} a => a ()) actions;
              wait (SOME (\kw{fn} () => List.all (\kw{fn} x => x) 
                                            (map isDone subgoals)));
              \kw{if} (List.all (\kw{fn} g => \kw{case} (goalStatus (g)) 
                                      \kw{of} Success => true 
                                       | _ => false) 
                           subgoals)
                \kw{then} (app goalClear subgoals;
                      goalSucceed (goal))
              \kw{else} (app goalClear subgoals;
                    goalFail (goal))
            \kw{end}\}
  \kw{end}
\end{code}
\hrule
\caption{Code for \expr{behaviorRule}}
\label{f:behaviorrule}
\end{figure}

\section{Implementation}
\label{s:impl}

The framework we have described has been implemented for the Standard
ML of New Jersey compiler \cite{Appel91} using the reactive library
described in \cite{Pucella98}. One advantage of this approach is that
the semantics of the system is easily derivable from the one in
\cite{Pucella98}. Before discussing the implementation, let us give an
overview of the reactive library.  

The library defines a type \expr{rexp} of reactive expressions, which
are expressions that define control points. A reactive expression is
created through a function \expr{rexp} that expects a \expr{unit
$\rightarrow$ unit} function as argument. The argument function calls
the function \expr{stop} to define a control point. The function \expr{react}
is used to take a reactive expression and to evaluate the code starting
from the last control point reached until the next control point is
reached. This is called \emph{activating} a reactive
expression. When a reactive expressions evaluates to a value without
reaching a control point, it is said to \emph{terminate}. Interesting
combinators can be defined to take reactive 
expressions and combining them. The most important of such combinators 
is the \expr{merge} combinator, which takes two reactive expression
$e_1$ and $e_2$ and creates a new reactive expression $e$ that behaves 
as follows: when $e$ is activated, $e_1$ and $e_2$ are activated,
one after the other. In effect, this interleaves the execution of
$e_1$ and $e_2$. The combinator \expr{loop} takes a reactive
expressions $e_1$ and creates a new reactive expression $e$ that
behaves as follows: when $e$ is activated, $e_1$ is activated. If
$e_1$ terminates, it is reset (the reactive expression is
re-instanciated) and activated again. The reactive expression
\expr{nothing} simply terminates immediately. 

A more fine-grained notion of control point is also
available. A reactive expression can call \expr{suspend} to suspend
its current execution. A \expr{suspend} acts as a \expr{stop}, except
that a special combinator \expr{close} is available. Given a reactive
expression $e_1$, \expr{close} returns a new reactive expression which 
behaves as follows: when $e$ is activated, it repeatedly activates
$e_1$ until all its reactive subexpressions have reached
{stop}-defined control points. This allows finer control over the
order of execution of the reactive subexpressions, an example of which 
we will see in this section. 

It is clear, given this description, how the library may be useful to
implement the details of  overlapping rules. For brevity, we assume
the reactive library has been  
bound to a structure \expr{R}. The implementation of the framework is
rather simple, although it is complicated by technical 
details and some mismatches with the underlying reactive library. We
define a rule set as a reactive expression, a \emph{merge} of all the
relevant rules.
\begin{code}
  \kw{type} rule_set = R.rexp
  \kw{fun} newSet () = R.nothing
\end{code}

A rule is defined as before, while adding a rule to a set of rules
consists of merging the reactive expression corresponding to the rule
in the merge of the rule set.
\begin{code}
  \kw{type} rule = \{cond : unit -> word, 
               action : unit -> unit\}

  \kw{fun} addRule rs \{cond,action\} = \kw{let}
    \kw{val} r = R.exp (\kw{fn} () => (condition (cond);
                             action ()))
  \kw{in}
    R.merge (rs,r)
  \kw{end}

  \kw{val} mkSet = foldr addRule (newSet ())
\end{code}

The function \expr{condition} terminates immediately if monitoring
determines that it should be fired (according to the fitness of the
condition), otherwise it stops to wait for another instant
where the condition is deemed fit. To bypass a limitation of the current reactive library,
which is more geared towards locally determining whether a given
reactive expression is allowed to continue rather than being selected
through a global check, we introduce a global variable to hold a list
of all computed fitnesses and allow the system to select the
ones that will execute \footnote{This does make the library non-reentrant. This could 
be corrected by an appropriate change to the reactive library
(implementing reaction-specific data, for instance), or by including a
notion of execution context to bind the use of the variable to a given 
context.}.
\begin{code}
  \kw{val} globalFitnesses = ref ([]) : (unit ref * word) list ref
\end{code}

We uniquely tag each condition being computed using a value of type
\expr{unit ref}, a trick commonly used in SML to get unique
identifiers that can be quickly compared for identity. When the
function \expr{condition} is encountered, the current fitness is 
computed and stored along with a unique identifier for the
condition. The reactive expression is then \emph{suspended} (not
stopped). This gives the other reactive expressions running \emph{in
parallel} a chance to evaluate their conditions. Upon resumption of the
suspension, each condition checks if it is allowed to continue by
seeing if it is listed in a list of \emph{allowed-to-continue}
conditions, stored in the previous \expr{globalFitnesses} variable.
\begin{code}
  \kw{fun} condition (f) = \kw{let}
    \kw{val} r = ref ()
    \kw{fun} loop () = (globalFitnesses := (r,f)::(!globalFitnesses)
                   suspend ();
                   \kw{if} (List.exists (\kw{fn} (r',_) => r=r') (!globalFitnesses))
                     \kw{then} ()
                   \kw{else} (stop (); loop ()))
  \kw{in}
    loop ()
  \kw{end}
\end{code}

The function \expr{wait} that implements an
interruption of the execution of the rule is simple:
\begin{code}
  \kw{fun} wait (NONE) = (stop (); suspend ())
    | wait (SOME (f)) = (stop (); condition (f))
\end{code}
(The suspend in \expr{wait (NONE)} is a technicality, needed to
prevent the firing of stopped rules with no conditions until all the
conditions of the other rules have been checked).

The core of the work is done in \expr{monitor}, which is in charge of activating the 
reactive expression corresponding to the rule set, gather the results, 
compute which conditions are satisfied, and resume the suspensions.
\begin{code}
  \kw{fun} monitor c rs = \kw{let}
    \kw{val} clearVar = R.loop (R.rexp (\kw{fn} () => (globalFitnesses := [];
                                              stop ())))
    \kw{val} r = R.loop (R.rexp (\kw{fn} () => (computeEnabled (c);
                                      stop ())))
  \kw{in}
    R.react (R.close (R.merge (clearVar,R.merge (rs,r))))
  \kw{end}
\end{code}

The above code encompasses the control flow of the monitoring process, 
and relies heavily on the fact that merges are deterministic. A
non-deterministic implementation could play with \expr{suspend} calls
to achieve the right order of execution: we need to make sure at every
instant that \expr{clearVar} is executed first, and
\expr{computeEnabled} is activated after all suspensions.
The actual rule selection is performed by \expr{computeEnabled}: 
\begin{code}
  \kw{fun} computeEnabled (AllBest) = \kw{let}
        \kw{fun} max (curr,[]) = curr
          | max (curr,(_,x)::xs) = \kw{if} x>curr \kw{then} max (x,xs) \kw{else} max (curr,xs)
        \kw{val} bestVal = max (1,!globalFitnesses)
        \kw{val} lst = List.filter (\kw{fn} (_,a) => a=bestVal) (!globalFitnesses)
      \kw{in} 
        globalFitnesses := lst
      \kw{end}
    | computeEnabled (RandBest) = \textit{(as AllBest, but return random element)}
    | computeEnabled (AllDownTo (v)) = \kw{let}
        \kw{val} lst = List.filter (\kw{fn} (_,a) => (a>=v)) (!globalFitnesses)
      \kw{in} 
        globalFitnesses := lst
      \kw{end}
    | computeEnable (RandDownTo (v)) = \textit{(as AllDownTo, but return random element)}
\end{code}

\section{Conclusion}

So what have we done? We have developed a general framework for
rule-based programming in SML, flexible enough to handle standard
OPS-style rules, as well as overlapping rules. The fact that rules are 
first-class in the framework gave us enough flexibility to
express Loyall's active behavior trees through an interpreter of
declarative behaviors. 

We have not worried about the efficiency of the framework. Some rather 
standard optimizations as found in modern rule-based systems can
easily be applied, although optimizations of the conditions may
require them to be lifted into a more structured type, such as for
example
\begin{code}
  \kw{datatype} condition = Basic \kw{of} unit -> word
                     | And \kw{of} condition list
                     | Or \kw{of} condition list
                     | Not \kw{of} condition
                     | ...
\end{code}
in order to allow for such things as caching of condition evaluations
and so on. On another note, evaluating a condition is only required
if the references the condition refers to have been changed, so an
optimized framework should maintain a list of the references used by
conditions and record changes accordingly. 

The tradeoff between generality and conciseness is not new. If we are
willing to restrict flexibility, we can design an appropriate surface language
which we can translate into this framework for execution. This is what 
we did for the implementation of behaviors in Section
\ref{s:abts}. This makes our framework a target language for the
interpretation of domain-specific rule-based languages. In a similar
way, the reactive library of \cite{Pucella98} was designed as a target 
language for the interpretation of domain-specific reactive
languages, which is the way it is being used in this paper. 

Using the SugarCubes framework described in \cite{Boussinot98}, we could
move most of the framework to Java, but SML has distinct advantages,
at least at first brush: we have seen how first-class rules and
first-class functions helped design suitable domain-specific
abstractions; the module system, although not used in this paper,
plays a crucial role in the generalization of the framework to general 
rules that can carry arbitrary data (not only conditions and
actions). The whole approach is also helped by the fact that SML
already implements state-based programming through the use of explicit
references. It will be interesting to see how much of this can be
carried over to Java. 

This frameworks, one of the first implemented using the library in
\cite{Pucella98}, also points to some conclusions about the reactive
approach. If the order in which parallel reactive expressions are
activated is important, then we need to be careful, appropriately
using \expr{suspend} calls to get the order right; this makes the 
resulting system brittle and the code hard to see correct. A better 
approach may be to allow one to explicitly control the order of
execution of the branches of a merge. On a related note, the reactive
library is geared towards determining reactive expression activation
locally, while we have encountered in this paper a reasonable instance 
where the activation decision is taken on a global level. It would be
interesting to find an extension of the reactive library to do that
cleanly, without resorting to a list of unique identifiers indicating
which expression is allowed to resume.

\COMMENTOUT{
\textbf{Future work.} Eventually, comparison with CML \cite{Reppy99},
in the style of \cite{Boussinot00}. Ensure actual reactivity. 
}

\bibliographystyle{plain}
\bibliography{main}

\end{document}